\documentclass[apl,twocolumn]{revtex4-1}%

\usepackage{graphicx}
\usepackage{dcolumn}
\usepackage{bm}
\newcommand{\Vec}[1]{\mbox{\boldmath$#1$}}

\begin{document}

\preprint{APS/123-QED}

\title{Corrugated flat band as an origin of large 
thermopower in hole doped PtSb$_2$}

\author{Kouta Mori$^{1,3}$}
\author{Hidetomo Usui$^{1,3}$}
\author{Hirofumi Sakakibara$^2$}
\author{Kazuhiko Kuroki$^{1,3}$}

\affiliation{$\rm ^1$Department of Engineering Science, The University of Electro-Communications, Chofu, Tokyo 182-8585, Japan}
\affiliation{$\rm ^2$Department of Applied Physics and Chemistry, The University of Electro-Communications, Chofu, Tokyo 182-8585, Japan}
\affiliation{$\rm ^3$ JST, ALCA, Sanbancho, Chiyoda, Tokyo 102-0075, Japan}

\date{\today}

\begin{abstract}
The origin of the recently discovered large thermopower in hole-doped PtSb$_2$ is theoretically analyzed based on a model constructed from first principles band calculation. It is 
found that the valence band dispersion has an overall flatness combined with some local ups and downs, which gives small Fermi surfaces  scattered over the entire Brillouin zone. The Seebeck coefficient is calculated using this model, which gives good agreement with the experiment. We conclude that the good 
thermoelectric property originates from this "corrugated flat band", 
where the coexistence of large Seebeck coefficient and large electric 
conductivity is generally expected.
\end{abstract}

\pacs{  }
\maketitle

Good thermoelectric materials are those materials that can transform 
heat into electricity with high efficiency.
The efficiency of thermoelectric materials is characterized by the 
dimensionless figure of merit, $ZT$, where $T$ is the 
temperature, and $Z=S^2\sigma/\kappa$ with $S$, $\sigma$ and $\kappa$ 
being the Seebeck coefficient, electric conductivity, and the 
thermal conductivity, respectively\cite{Mahanrev}. 
In particular, the product $P=S^2\sigma$ 
is called the power factor, and a material with large $P$ requires a 
large $S$ and $\sigma$. It is known, however, that 
materials with large Seebeck coefficient usually have small conductivity, 
so that good thermoelectric materials are often found in semiconducting 
materials with rather small amount of carriers. 

\begin{figure}[!b]
\includegraphics[width=9cm]{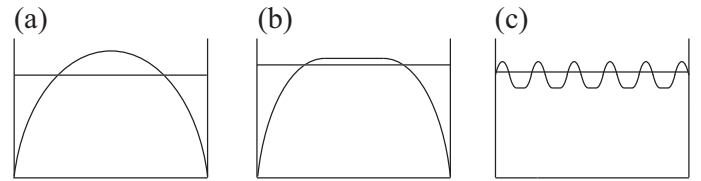}
\caption{A schematic image of (a) usual metal band 
(b) pudding mold band (c) corrugated flat band. The horizontal lines
are image of the position of the 
Fermi level that gives the same Fermi surface volume.
}
\label{fig1}
\end{figure}

In this context, the discovery of 
good thermoelectric properties in the sodium 
cobaltate Na$_x$CoO$_2$\cite{Terasaki} has been 
of special interest in that a large power factor is observed in a material with 
large amount of doped holes and thus metallic conductivity. 
There have been various theoretical studies on this material
\cite{Singh,Koshibae,KoshibaePRL,Kuroki}, and in particular, one of the 
present authors along with Arita proposed that a peculiar band shape named the 
``pudding mold type'' (Fig.\ref{fig1}(b)) 
is the origin of this coexistence of good conductivity 
and large thermopower\cite{Kuroki}. Namely in a system having a 
band with flat portion at the top (or bottom), 
connecting into a dispersive portion, the Fermi level is kept 
close to the band edge upon doping, 
and this gives rise to a large group velocity 
difference between electrons and holes, 
resulting in a large Seebeck coefficient.
At the same time, the large group velocity of holes due to the 
dispersive portion of the band gives a large electric conductivity, 
and this combination results in a large power factor.

Quite recently, Nishikubo {\it et al.} discovered that doping holes into 
a cubic pyrite material PtSb$_2$ by partially substituting 
Pt by Ir as Pt$_{1-x}$Ir$_x$Sb$_2$ gives rise to a coexistence of metallic 
conductivity and a large Seebeck coefficient\cite{Nohara}.
Refering to the band structure calculation in ref.\cite{Emtage}, 
a possible relevance of the peculiar band structure has been pointed out.
In the present paper, we present first principles band calculation 
result of PtSb$_2$, and analyze the origin of the good thermoelectric 
properties of this material. It is shown that the material has a 
valence band dispersion 
with an overall flatness but with some local ups and downs. 
This band structure, which we refer to as the ``corrugated flat band'' 
(Fig.\ref{fig1}(c)),  
has some similarity with the pudding mold band in that they both have 
coexisting flatness and the large group velocity, but still is different 
from the pudding mold band in that the flatness extends over the entire 
Brillouin zone,  while the dispersiveness (the corrugation) is a local 
feature which appears at various points all over the Brillouin zone.

\begin{figure}[!b]
\includegraphics[width=8cm]{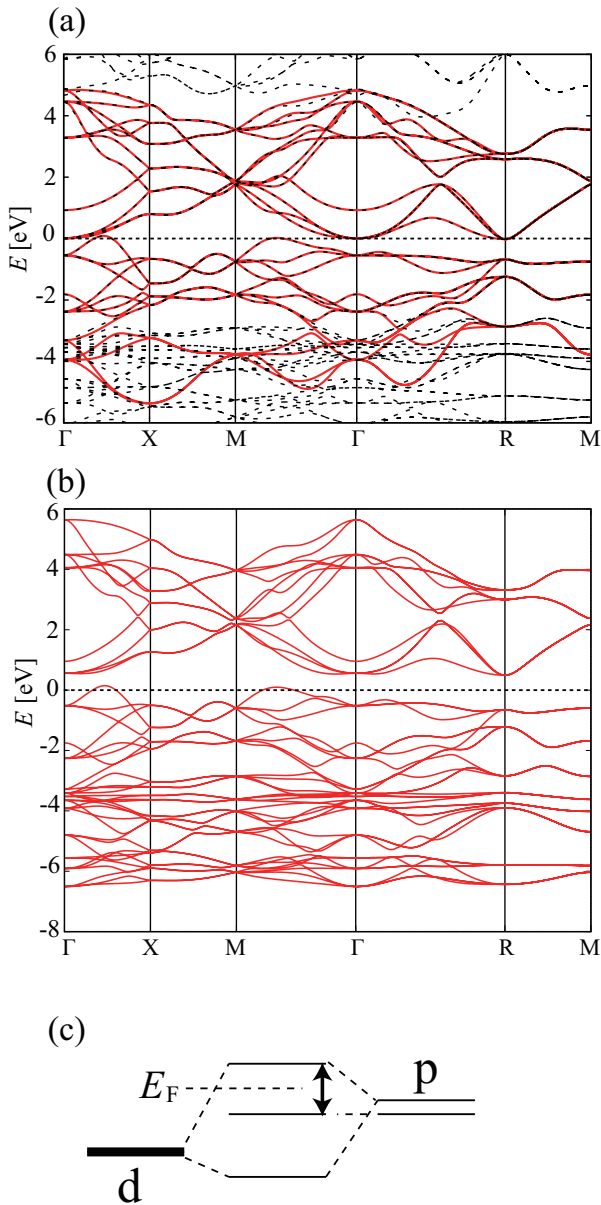}
\caption{(a) Dashed lines : the first principles band calculation 
result of PtSb$_2$. Red solid lines : 
the band structure of the 24 orbital model.
(b) The band structure of 44 orbital $p$-$d$ model with all of the 
$d$-$p$ hybridization multiplied by a factor of 1.2.
(c) A schematic figure of the $p$-$d$ hybridization and the band gap.}
\label{fig2}
\end{figure}   

\begin{figure}[!t]
\includegraphics[width=8cm]{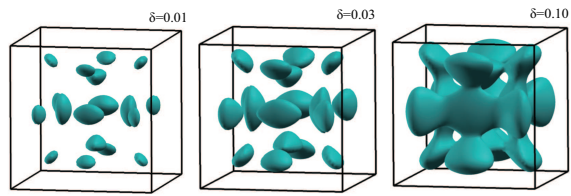}
\caption{The Fermi surface of the 24 band model for 
(a) $\delta=0.01$ (b) 0.03 (c) 0.1. }
\label{fig3}
\end{figure}   

First principles band calculation of PtSb$_2$ is performed 
using the Wien2K package\cite{Wien2k}. We use the experimentally 
determined lattice structure given in ref.\cite{structure} 
The spin-orbit coupling, which turns out to have only small effect, 
is omitted in the present results.
The calculation result is shown in Fig.\ref{fig2}(a) (dashed lines). 
The bands near the Fermi level mainly consists of Sb $5p$ orbitals, 
mixed with Pt $5d$ orbitals.
The upper most valence band originates almost solely from 
Sb $5p$ orbitals, and the near absence of the $d$ orbital 
character, namely, the non-bonding nature (non-bonding in the sense 
that $p$ does not mix with $d$),  makes this band 
relatively flat compared to other bands with $5p$ character. 
On the other hand, 
this upper most valence band is not perfectly flat, and 
there are some ups and downs, which we will call ``corrugations''. 
This corrugation 
give rise to small pocket like Fermi surfaces when holes are doped
as we shall see below. 

\begin{figure}[!t]
\includegraphics[width=7.5cm]{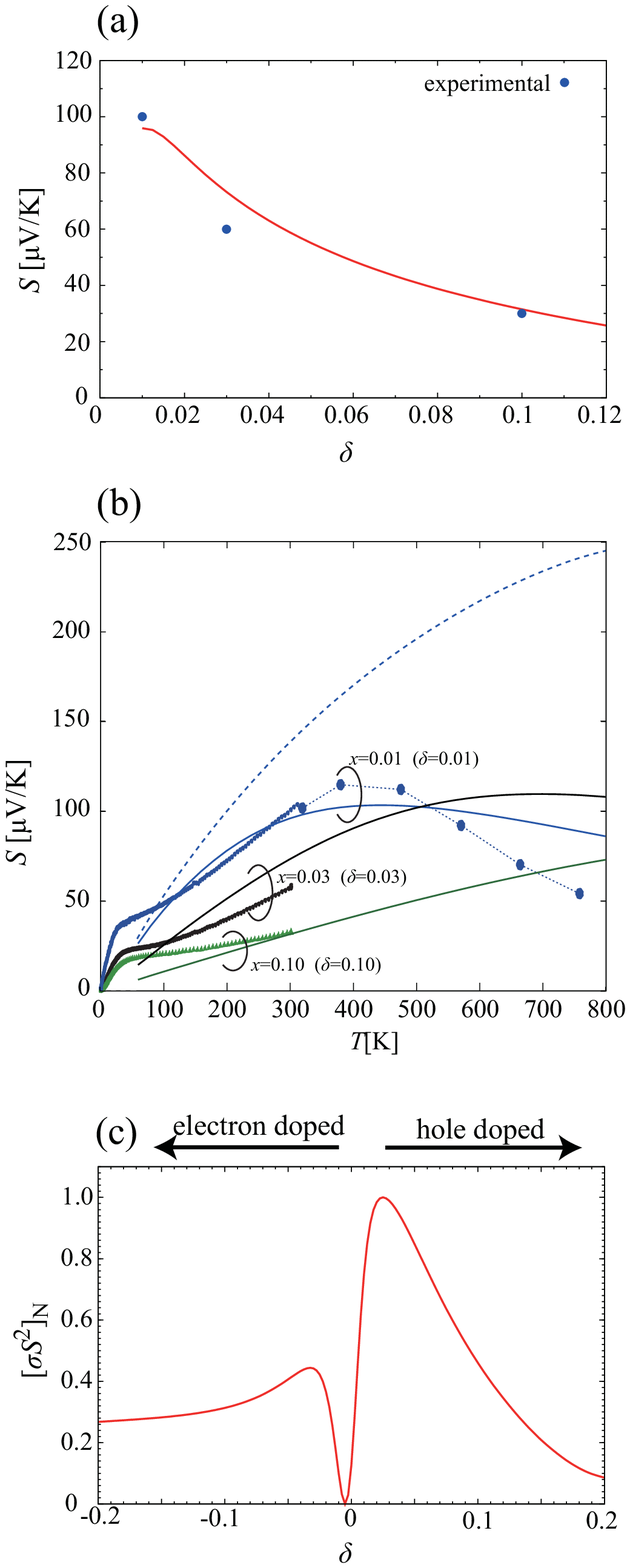}
\caption{(a)The calculated Seebeck coefficient against the doping 
ratio $\delta$ at $T=300$K. The dots are the experimental data from 
ref.\cite{Nohara} (b) The Seebeck coefficient against the 
temperature for various doping ratios. (Connected) symbols are the 
experimental data from ref.\cite{Nohara}. Upper most dashed line 
is the result for the 
hypothetical band structure shown in Fig.\ref{fig2}(b) with $\delta=0.01$. 
(c) The power factor 
(normalized with its maximum value) against the doping ratio for $T=300$K.
}
\label{fig4}
\end{figure}

From this band calculation, we obtain maximally localized 
Wannier orbitals\cite{Wannier,w2w} mainly consisting of Sb $5p$ orbitals, 
which enables us to 
construct a 24 orbital (eight Sb per unit cell $\times$ three $5p$  
orbitals) tightbinding model that correctly reproduces the 
original first principles band structure around the Fermi level
(Fig.\ref{fig2}(a), solid lines).
We consider doping holes into this model assuming a rigid band, 
where the doping ratio $\delta $ is defined as $\delta=(24-n)/4$ with 
$n$ being the total band filling. 
In an ideal situation, this $\delta$ corresponds to the experimental 
substitution ratio $x$.
In Fig.\ref{fig3}, we show the Fermi surface for 
three doping ratios.
The combination of the overall flatness and the corrugation 
of the valence band 
results in multiple Fermi surface pockets that are scattered over the entire
Brillouin zone.

Based on this model, we calculate the Seebeck coefficient using the 
Boltzmann's equation approach. 
\begin{equation}
{\bf S}=\frac{1}{eT}{\bf K}_0^{-1}{\bf K}_1
\end{equation}
where $e(<0)$ is the electron charge, $T$ is the temperature, 
tensors ${\bf K}_0$ and ${\bf K}_1$ are given by
\begin{equation}
{\bf K}_n=\sum_{\Vec{k}}\tau\Vec{v}(\Vec{k})\Vec{v}(\Vec{k})
\left[-\frac{\partial f(\varepsilon)}
{\partial \varepsilon}(\Vec{k})\right]
(\varepsilon(\Vec{k})-\mu)^n.
\label{eq2}
\end{equation}
Here, $\varepsilon(\Vec{k})$ is the band dispersion of the tight binding model, 
$\Vec{v}(\Vec{k})=\nabla_{\Vec{k}}\varepsilon(\Vec{k})$ is the 
group velocity, $\tau$ is the quasiparticle lifetime, which will be 
taken as a constant in the present study. 
$f(\varepsilon)$ is the Fermi distribution function,  
and $\mu$ is the chemical potential determined from the band filling. 
Hereafter, we simply refer to $({\bf K}_n)_{xx}$ as $K_n$, and 
$S_{xx}=(1/eT)\dot(K_1/K_0)$ (for diagonal ${\bf K}_0$) as $S$. 
Using $K_0$, conductivity can be given as 
$\sigma_{xx}=e^2K_0\equiv\sigma={1/\rho}$.

In Fig.\ref{fig4}(a), we show the doping ratio dependence of the Seebeck 
coefficient at $T=300$K. It can be seen that the calculation results are 
in overall agreement with the experiment\cite{comment}.
In Fig.\ref{fig4}(b), we show the temperature dependence of the 
Seebeck coefficient. This is also in 
fair agreement with the experiment, although the steep increase at 
very low temperatures seen experimentally 
is not found in the calculation. In particular, 
the broad maximum at around $T=400$K observed for $x=0.01$ 
is roughly reproduced theoretically. This maximum 
is due to the presence of the conduction band lying close to the 
Fermi level, which gives 
a negative contribution to the Seebeck coefficient at high temperatures.
We will come back to this point later.

In Fig.\ref{fig4}(c), we show the normalized power factor as a function of the 
doping. It can clearly be seen that a large power factor arises in the 
hole doped regime, where the Fermi level lies in the corrugated flat band.
This good thermoelectric property can be understood as follows. 
In metallic systems, where the Fermi level lies within the energy bands, 
it is generally required to keep the Fermi level close to the band edge 
in order to obtain a large Seebeck coefficient. This is because 
the Seebeck effect is caused by the difference of the 
group velocity between electrons and holes in the vicinity of the 
Fermi level, and the two velocities differ largely (the ratio of the 
group velocities above and below the Fermi level is large) near the band edge.
However, keeping the Fermi  level close to the band edge usually 
means that the number of doped carriers is small, so that the absolute values
of the group velocity and the volume of the Fermi surface is both small,
resulting in a small electric conductivity.
Similar to the case of the pudding mold band\cite{Kuroki}, the situation 
is largely different in the case of the corrugated flat band.
As mentioned above, multiple 
Fermi pockets are scattered over the entire Brillouin zone due to the 
flatness of the conduction band. This situation is schematically shown in 
Fig.\ref{fig1}(c) along with the case of a usual metal 
with a single Fermi surface Fig.\ref{fig1}(a).
The overall flatness of the bands and thus the large multiplicity of the 
Fermi surface prevents the Fermi level from quickly lowering upon 
hole doping, giving rise to a large Seebeck coefficient even  for large 
amount of hole doping. At the same time, 
the existence of multiple Fermi surfaces together with a large 
group velocity due to the corrugation results in a large electric conductivity.
The combination of these effects results in a large power factor.
In other words, in a corrugated flat band,  
the  multi-valley effect seen in some of the 
thermoelectric semiconducting materials\cite{Mahanrev} is enhanced in 
an ultimate way in that the overall flatness of the band extends over the 
entire Brillouin zone.

As seen from the above, the corrugated flat band  is indeed 
a favorable band structure for large thermopower, 
but there exists a flaw in the 
present material in that the conduction band lies too close to the valence band.
The conduction band gives a negative contribution, 
so that the Seebeck coefficient decreases at high temperatures.
Since the band gap opens between the nonbonding and the antibonding $p$-$d$ 
bands as shown in Fig.\ref{fig2}(c),
a larger band gap is expected to open for larger $p$-$d$ 
hybridization, where the bonding-antibonding splitting becomes large.
To see this effect, we construct a 44 orbital model which explicitly considers 
20 (5 orbitals $\times$ 4 sites) Pt $5d$ and 
24 Sb $5p$ orbitals, and increase all of the hopping integrals 
between $p$ and $d$ orbitals by a factor of 1.2 hypothetically by hand. 
The obtained band structure is shown in Fig.\ref{fig2}(b).
For such a band structure, the Seebeck coefficient continues to increase 
for high temperatures as shown in Fig.\ref{fig3}(b) (dashed line). 

To summarize, we have shown that the recently found good 
thermoelectric properties of PtSb$_2$ is due to the presence of 
the corrugated flat band, which gives rise to multiple Fermi pockets 
scattered over the entire Brillouin zone. The flatness of the 
band originates from the non-bonding character of the Sb $p$ band,
while the corrugation of this band mainly comes from the direct $p$-$p$ 
hoppings. The multiplicity of the 
Fermi surface results in a coexistence of large Seebeck coefficient and 
a good metallic conductivity, giving rise to a large power factor.
The present mechanism for large power factor 
is rather general, and is expected to 
take place where bands with non-bonding character is present near the 
Fermi level. Despite the presence of the corrugated flat band, 
one flaw of PtSb$_2$ is the narrowness of the band gap.
The conduction band lying close to the valence band gives a 
negative conductivity to the Seebeck coefficient at high temperatures.
We expect that materials with the combination of a corrugated flat band along 
with a wider band gap 
should result in even more ideal thermoelectric properties.

\section*{ACKNOWLEDGMENTS}

We are grateful to M. Nohara for showing us the experimental 
results prior to publication. H.S. acknowledges support from JSPS.


\begin{thebibliography}{99}
\bibitem{Mahanrev} For a general review on the theoretical 
aspects as well as experimental observations 
of thermopower, see,  G.D. Mahan {\it Good Thermoelectrics, 
Solid State Physics {\bf 51}, 81 (1997).}
\bibitem{Terasaki}
I. Terasaki, Y. Sasago and K. Uchinokura, Phys. Rev. B {\bf 56} R12685 (1997).
\bibitem{Singh}
D.J. Singh, Phys. Rev. B {\bf 61}, 13397 (2000).
\bibitem{Koshibae}
W. Koshibae, K. Tsutsui and S. Maekawa, Phys. Rev. B {\bf 62} 6869 (2000).
\bibitem{KoshibaePRL}  W. Koshibae and S. Maekawa, Phys. Rev. Lett. {\bf 87}, 
236603 (2001).
\bibitem{Kuroki}
K. Kuroki and R. Arita, J. Phys. Soc. Jpn. {\bf 76} 083707 (2007).
\bibitem{Nohara} 
Y. Nishikubo, S. Nakano, K. Kudo, and M. Nohara, Appl. Phys. Lett. {\bf 100},
252104 (2012).
\bibitem{Emtage} P.R. Emtage, Phys. Rev. B {\bf 138}, A246 (1965).
\bibitem{Wien2k}
P. Blaha, K. Schwarz, G.K.H. Madsen, D. Kvasnicka, and J. Luitz, 
{\it Wien2k: An Augmented Plane Wave} + {\it Local Orbitals Program for Calculating Crystal Properties} (Vienna University of Technology, Wien, 2001).
Here we take $RK_{\rm max}=10$, 1024 $k$-points, and 
adopt the exchange correlation functional introduced by 
J. P. Perdew , K. Burke, and M. Ernzerhof 
[Phys. Rev. Lett. {\bf 77}, 3865 (1996)].
\bibitem{structure} N.E. Brese and H.G. Von Schnering, Z. Anorg. Allg. Chem., 
{\bf 620}, 393 (1994).
\bibitem{Wannier} N. Marzari and D. Vanderbilt, Phys. Rev. B 
{\bf 56}, 12847 (1997);  
I. Souza, N. Marzari, and D. Vanderbilt, Phys. Rev. B {\bf 65}, 035109 (2001).
The Wannier functions are generated by the code developed by
A. A. Mostofi, J. R. Yates, N. Marzari, I. Souza, and D. Vanderbilt,
(http://www.wannier.org/).  
\bibitem{w2w} J. Kunes, R. Arita, P. Wissgott, A. Toschi, H. Ikeda, and K. Held, Comp. Phys. Commun. {\bf 181} 1888 (2010).
\bibitem{comment} Here we omit the calculation result for the non-doped 
case. This is because the Seebeck coefficient for $\delta=0$ 
is very sensitive to the band gap, which cannot be estimated with 
enough accuracy in DFT calculations, especially when the gap is small.
\end{thebibliography}
\end{document}